\documentclass[%
 reprint,
 amsmath,amssymb,
 aps,
]{revtex4-2}

\usepackage{graphicx}
\usepackage{dcolumn}
\usepackage{bm}
\usepackage{braket}

\usepackage{hyperref}
\hypersetup{
    colorlinks=true,
    linkcolor=blue,
    filecolor=magenta,
    urlcolor=blue,
    citecolor=blue,
    bookmarks=true,
    pdfpagemode=FullScreen,
}

\begin{document}

\preprint{APS/123-QED}

\title{Amorphization-induced topological and insulator-metal transitions in bidimensional Bi$_x$Sb$_{1-x}$ alloys}

\author{A. J. Ur\'ia-\'Alvarez}
\email{alejandro.uria@uam.es}
\affiliation{Departamento de F\'isica de la Materia Condensada, Condensed Matter Physics Center (IFIMAC), and Instituto Nicol\'as Cabrera (INC),
Universidad Aut\'onoma de Madrid, Cantoblanco 28049, Spain}

\author{J. J. Palacios}
\email{juanjose.palacios@uam.es}
\affiliation{Departamento de F\'isica de la Materia Condensada, Condensed Matter Physics Center (IFIMAC), and Instituto Nicol\'as Cabrera (INC),
Universidad Aut\'onoma de Madrid, Cantoblanco 28049, Spain}

\date{\today}

\begin{abstract}
Bismuth has been shown to be topological in its different allotropes and compounds, with one of the most notable examples being the Bi-Sb alloy, the first 3D topological insulator ever discovered. In this paper we explore two-dimensional alloys of Bi and Sb, both crystalline and amorphous, to determine the critical concentrations that render the alloys topological. For the amorphous alloy, we determine the effect of structural disorder on its topological properties, remarkably observing a trivial to topological transition as disorder increases. The alloys are modelled using a Slater-Koster tight-binding model and the topological behaviour is assessed through the entanglement spectrum together with artificial neural networks. Additionally, we perform electronic transport calculations with results compatible with those of the entanglement spectrum, which, furthermore, reveal an insulator to metal transition in the highly disordered regime.
\end{abstract}

\maketitle


\section{Introduction}

With the advent of topological phases of matter \cite{haldane1988model, kane2005}, significant effort has been devoted to the characterization of these phases, particularly into the identification of quantities that distinguish these materials, i.e. topological invariants and markers. Multiple approaches have been developed, being one of the most successful the Wilson loop \cite{soluyanov2011computing, alexandradinata2014wilson, gresch2017z2pack}, which allows to extract either the Chern number \cite{thouless1982quantized} or the $\mathbb{Z}_2$ invariant \cite{kane2005z}, and can also be used to characterize Weyl semimetals \cite{weng2015weyl, saini2022wloopphi} or higher-order topological insulators (HOTIs) \cite{benalcazar2017quantized, franca2018anomalous}. This method enables high-throughput calculations to unveil the topology of families of materials \cite{olsen2019discovering, mounet2018two}. However, it is with the topological quantum chemistry approach \cite{bradlyn2017topological} that the topological classification of crystalline materials is becoming complete \cite{vergniory2019complete}. Consequently, the focus is shifting to disordered phases, i.e. systems where the common techniques to compute topological invariants, such as the Wilson loop, become less effective or entirely fail due to the absence of a gap in the system.

Several quantities have been developed to determine the topology of disordered systems, namely those lacking translational symmetry. For Chern insulators, notable examples are the Chern marker \cite{bianco2011mapping, ceresoli2007} and the Bott index \cite{loring2011disordered, hastings2011topological}. For time-reversal topological insulators (TIs), examples include the spin Chern marker \cite{prodan2009, Favata_2023, bau2024}, spin Bott index \cite{huang2018quantum, huang2018theory} or the Pfaffian-Bott index\cite{loring2011disordered}. More recently, the spectral localizer \cite{cerjan2022local, franca2024topological} has been introduced. Other measures that reflect topological features without directly computing the system's invariant include the spillage \cite{munoz2023structural}, which uses the projector onto the ground state to detect band inversions, and the entanglement spectrum \cite{li2008entanglement, prodan2010entanglement, alexandradinata2011trace, brzezinska2018entanglement, turner2010entanglement, fidkowski2010entanglement, hughes2011inversion, turner2009band, mondragon2013characterizing, legner2013relating, wang2014entanglement, budich2014topological, lee2015free, huang2012entanglement, lee2014position}, where a spatial cut of the ground state can reveal topological edge states. Machine learning (ML) techniques, especially artificial neural networks (ANNs), can also successfully determine topological invariants. By inputting physical data such as the density matrix \cite{carvalho2018real, che2020topological}, Hamiltonian \cite{zhang2018machine, sun2018deep}, wavefunctions \cite{mano2019application, holanda2020machine, scheurer2020unsupervised} or Berry curvatures \cite{molignini2021supervised}, ANNs can infer the invariant. Overall, ANN have found extensive applications in condensed matter physics \cite{carrasquilla2020, Bedolla_2021} and have kept up with the more modern developments of deep learning, such as attention mechanisms and transformers \cite{li2024deeplearningapproachsearch, viteritti2023, luo2022}.

\begin{figure}[b]
    \centering
    \includegraphics[width=1\columnwidth]{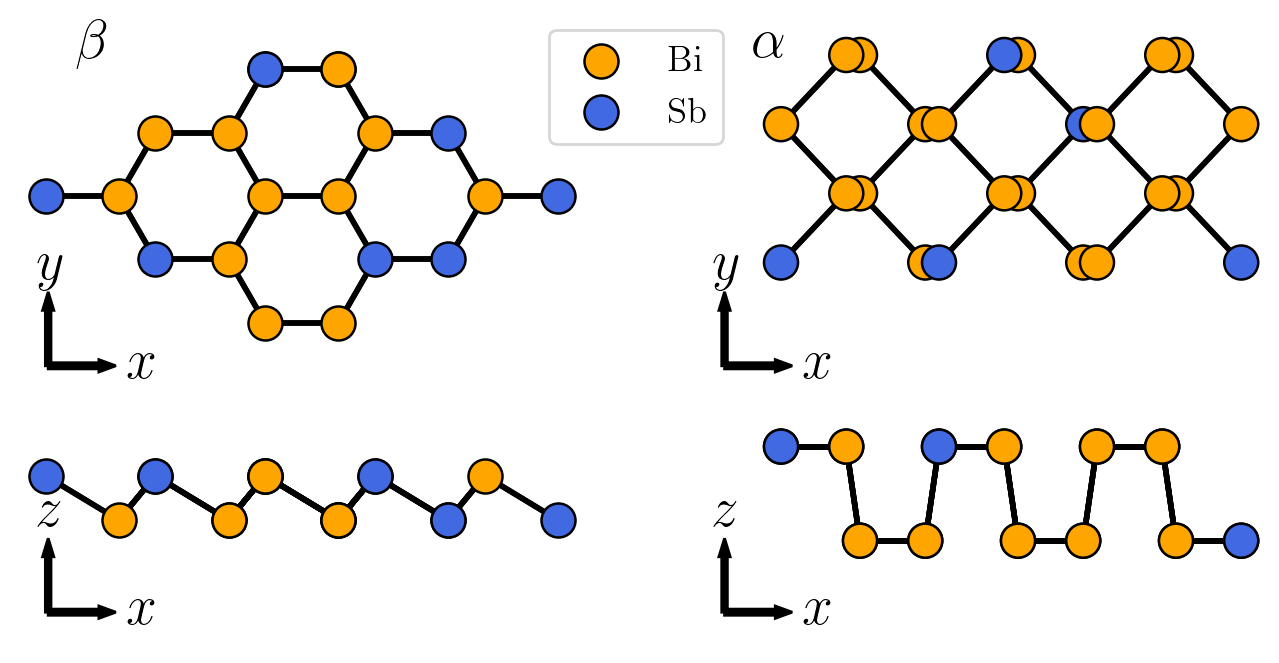}
    \caption{Two different crystalline allotropes of $\text{Bi}_x\text{Sb}_{1-x}$ alloys: the $\beta$ crystal, which corresponds to a buckled honeycomb lattice, and the $\alpha$ crystal for the puckered lattice.}
    \label{fig:crystal-alloys}
\end{figure}

Among disordered systems, alloys and amorphous systems hold particular significance due to their experimental accessibility and presence in multiple technologies \cite{Corbae_2023}. Only recently their topological properties began to be explored \cite{grushin2023topological, agarwala2019topological, yang2019topological, marsal2020topological, yuliang2023}. One notable group of materials are Bi-based compounds, which are known to be topological in crystalline form. For instance, Bi(111), $\beta$-bismuthene and $\alpha$-bismuthene exhibit topological behaviour in crystalline and amorphous states \cite{murakami2006, liu2011stable, costa2019toward, sabater2013topologically, syperek2022observation, bai2022}, bulk Bi is a HOTI \cite{schindler2018higher} and the three-dimensional alloy Bi$_x$Sb$_{1-x}$ is a strong topological insulator \cite{teo2008surface, hsieh2009observation}. This motivates us to study two-dimensional alloys of Bi$_x$Sb$_{1-x}$, in both crystalline and amorphous form. While the topological properties of the crystalline 2D alloy have already been addressed in previous works \cite{brzezinska2018entanglement, nouri2018topological, corbae2024hybridization}, the amorphous form remains unexplored.

In a previous study \cite{uria2022}, we combined the entanglement spectrum with ANNs to estimate the $\mathbb{Z}_2$ invariant in a simple model of TI. In the present work, we extend this methodology to map the topological phase diagram of the $\beta$ alloy (buckled lattice) and also explore the $\alpha$ lattice (puckered), typically overlooked for Bi. Furthermore, we examine the phase diagram as a function of the structural disorder, revealing a trivial to topological transition as disorder increases for a fixed concentration. This finding confirms a result that was previously reported for stanane \cite{wang2022}. With the aid of electronic transport calculations, we investigate the topological phase diagram in more depth and uncover an insulator-metal transition in the highly disordered alloy, which may reflect the behaviour of 3D amorphous Bi, known to be superconducting \cite{mata2016superconductivity}.

\section{Methodology}

\begin{figure*}
    \centering
    \includegraphics[width=1\textwidth]{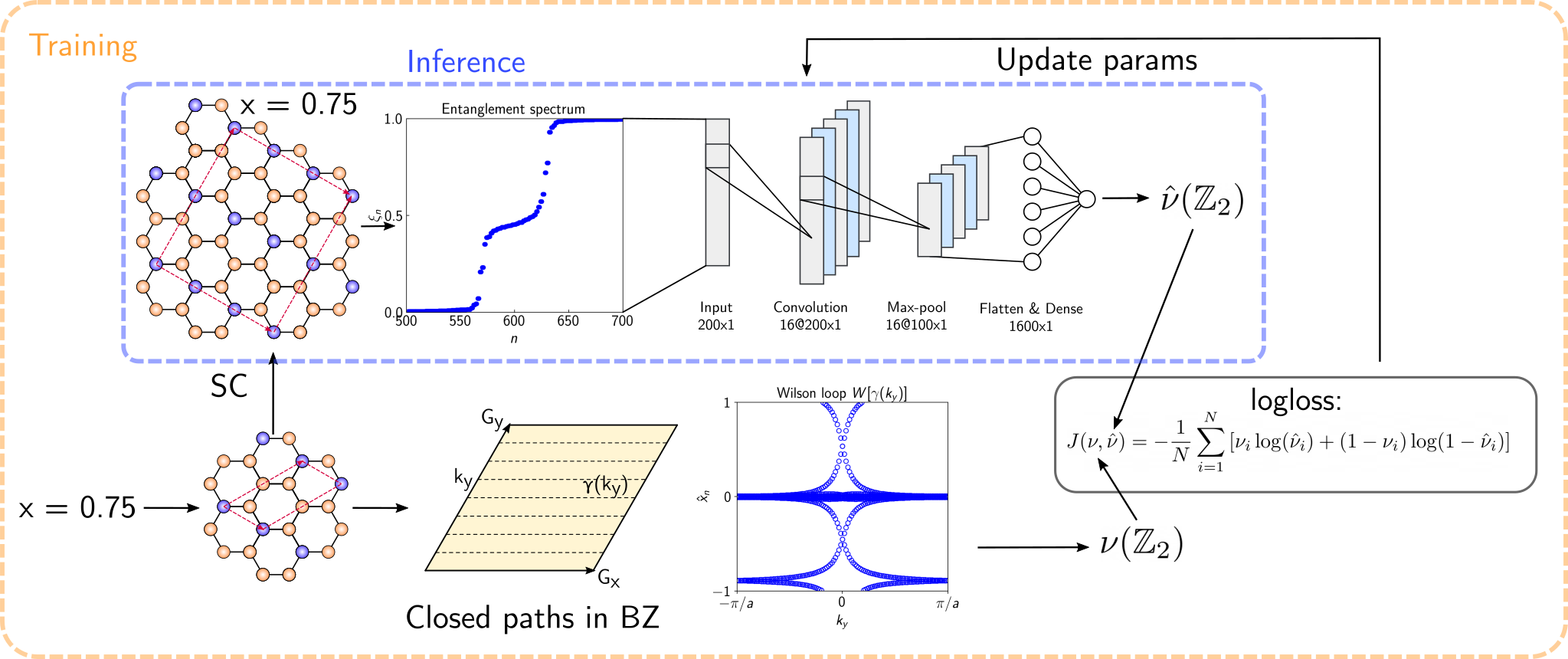}
    \caption{Description of the workflow with the artificial neural network. For training, we consider a small cell matching some concentration. We compute its invariant with the Wilson loop, and then extract the entanglement spectrum from a supercell. The neural network is trained using the (entanglement, invariant) pair. For inference, we consider a supercell for one specific concentration, obtain its entanglement spectrum and feed it to the ANN, which will predict its invariant.}
    \label{fig:nn_training}
\end{figure*}

Bi and Sb are described each one individually by a Slater-Koster (SK) tight-binding model, with the onsite energies, hopping amplitudes and spin-orbit coupling (SOC) taken from Ref. \onlinecite{liu1995} (see Appendix \ref{sec:sk_parameters}). These parameters, while originally devised for the three-dimensional semimetals, also reproduce correctly the band structure of the 2D compounds, as we will see later. In particular, for Bi it captures the topological behaviour of Bi(111) monolayers, which exhibit a quantum spin Hall state. The Hamiltonian of the tight-binding models is written in general as
\begin{align}
   \nonumber  H &= \sum_{i\alpha}\varepsilon_{i\alpha}c^{\dagger}_{i\alpha}c_{i\alpha} + \sum_{i\alpha,j\beta}t^{\alpha\beta}_{ij}c^{\dagger}_{i\alpha}c_{j\beta} \\
    &+ \lambda\sum_{i\alpha,j\beta}\braket{i\alpha|\mathbf{L}\cdot\mathbf{S}|j\beta}c^{\dagger}_{i\alpha}c_{j\beta}
\end{align}
where the indices $i,j$ run over lattice positions, and $\alpha,\beta$ run over orbital degrees of freedom. $\lambda$ denotes the strength of the SOC.
We use the same set of parameters to describe the multiple allotropes of Bi or Sb. For the alloys, we simply mix the hopping amplitudes whenever there is a hopping involving atoms of different species, i.e.
\begin{equation}
    t^{\alpha\beta}_{\text{Bi-Sb}} = \frac{1}{2}\left(t^{\alpha\beta}_{\text{Bi}} + t^{\alpha\beta}_{\text{Sb}}\right), \quad \forall\alpha,\beta
\end{equation}
Alternatively, the alloy can be modelled using the virtual crystal approximation (VCA), where we modify all the tight-binding parameters in a continuous way from pure Bi to pure Sb \cite{brzezinska2018entanglement}:
\begin{align}
    \nonumber \varepsilon^{\alpha}_x &= x \varepsilon^{\alpha}_{\text{Bi}} + (1-x)\varepsilon^{\alpha}_{\text{Sb}} \\
     t^{\alpha\beta}_{x} &= x t^{\alpha\beta}_{\text{Bi}} + (1-x)t^{\alpha\beta}_{\text{Sb}},\quad x\in[0,1],\quad \forall\alpha,\beta \\
    \nonumber \lambda_x &= x \lambda_{\text{Bi}} + (1-x)\lambda_{\text{Sb}}
\end{align}
where $x$ denotes the concentration of the alloy Bi$_x$Sb$_{1-x}$.

In the VCA note that we also have to modify the onsite energies and the SOC strength, while in the mixing approach we only change the hopping amplitudes, leaving the corresponding onsite energies and spin-orbit strength of the chemical species.
On this work we focus on three different allotropes of the Bi$_x$Sb$_{1-x}$ alloys: the crystalline $\alpha$ and $\beta$ forms, and amorphous forms. The $\alpha$ allotrope corresponds to a puckered lattice, whereas the $\beta$ one is buckled, as shown in Fig. \ref{fig:crystal-alloys}. The lattice parameters for the two crystals have been taken from \cite{akturk2016, nouri2018topological}, and we use the same for both Bi and Sb. For the crystalline alloys, we will use the mixing approach to obtain the interspecies hopping amplitudes. However, for the amorphous lattice we use a reduced version of the VCA to simplify the generation of the structures, where the hopping parameters are fixed and only SOC changes. The generation of the amorphous structure will be based for simplicity on the $\beta$ structure. Starting from the crystalline position, we introduce structural disorder via random displacements of the atomic positions. The magnitude of the displacement, $|\mathbf{\delta r}|$, is sampled from a uniform distribution:
\begin{align}
    \nonumber&\mathbf{r}' = \mathbf{r} + \mathbf{\delta r},\quad  \text{where}\\
    &|\mathbf{\delta r}|\sim U(0,\sigma r_0),\ \theta\sim U(0,2\pi),\ \varphi\sim U(0,\pi)
\end{align}
$\theta$, $\varphi$ being the angles of the displacement vector $\mathbf{\delta r}$ in spherical coordinates. $\sigma$ is the maximum displacement possible, i.e. the disorder strength, given in terms of the reference length (first neighbours in the crystalline case). Since the distances between the atoms will change as we increase $\sigma$, we need to modify the hopping parameters accordingly to properly capture the amorphous lattice. We introduce this dependency on the distance via an exponential law:
\begin{equation}
    t'^{\alpha\beta}(\mathbf{r})= t^{\alpha\beta}e^{-C(r-r_0)}\theta(R_c-r),\quad\forall\alpha,\beta
    \label{eq:hoppings_amorphous}
\end{equation}
where $r=|\mathbf{r}|$ and $C$ is the inverse decay length, $C=1$. Note that we have also introduced a cutoff distance $R_c$ via a Heaviside step function $\theta(r)$, which we set to $R_c=1.4r_0$. Thus, in the amorphous case, instead of defining a set of hopping parameters up to the $n$-th neighbour, the hopping parameters are defined up to a maximum distance, based on those to first neighbours in the crystal.

To assess the topological nature of all the structures considered, we use a combination of three different techniques. For the crystalline structures, we use the Wilson loop, which is defined as $W(k_y)=\prod_{i\in\gamma}\left(\sum_n\ket{u_{n\mathbf{k}}}\bra{u_{n\mathbf{k}}}\right)$, where $\gamma$ denotes a closed path in the Brillouin zone (BZ), and $\mathbf{k}=(k_i,k_y)$ \cite{vanderbilt2018berry}. It can be shown that its eigenvalues correspond to hybrid Wannier charge centers (HWCC) \cite{alexandradinata2014wilson}. By tracking the evolution of these centers as a function of $k_y$, one can extract the topological invariant of the system, in this case the $\mathbb{Z}_2$ invariant \cite{soluyanov2011computing}. Another quantity that reflects topological properties is the entanglement spectrum. By considering a spatial partition of the system, typically into two halves $A$ and $B$, one can construct the correlation matrix $C_{ij}=\braket{\Psi|c^{\dagger}_ic_j|\Psi}$, with $i,j\in A$. From the spectrum of this matrix, named the single-particle entanglement spectrum, we can extract the $\mathbb{Z}_2$ invariant \cite{alexandradinata2011trace}, and it was shown that one can obtain the entanglement spectrum for the actual many-body ground state $\ket{\Psi}$, which is the Fermi sea \cite{peschel2003calculation}.

The Wilson loop can be used in principle always as long as the system presents a global gap in the BZ. However, as the unit cell is enlarged and more bands appear in the system, tracking the HWCC becomes more difficult, on top of the computational expense it supposes to diagonalize a large Hamiltonian at each point of the BZ. For this reason, we use the entanglement spectrum as a proxy for the topological invariant: We compute the entanglement spectrum only at $\mathbf{k}=0$ and feed it to a ANN trained to identify the topological invariant from this spectrum. To train the neural network, we create a dataset formed by pairs of the actual invariant, computed by means of the Wilson loop, and the corresponding entanglement spectrum for a supercell. We do so considering different stoichiometries for both crystalline structures, as shown in Fig. \ref{fig:nn_training}. Then, varying the SOC we can generate enough data points for both the trivial and topological regions, for all the different stoichiometries. For the amorphous lattice we proceed analogously: we generate data from the crystalline case, changing SOC, and also train the network with data from the disordered case, where we know that the gap will still be open.

To verify the predictions of the neural network, we look for the standard signatures of a topological phase, which is the presence of edge states, either by looking at the local density of states, or by computing the conductance of a finite ribbon, using the Landauer-Buttiker formalism. The conductance is given in terms of the transmission $T(E)$, which is computed using the Caroli formula (see Appendix \ref{app:transport}). If topological edge states are present, we expect to observe a quantized conductance at the Fermi energy.

\section{Results}

\subsection{$\alpha$- and $\beta$-Bi$_x$Sb$_{1-x}$}

\begin{figure}
    \centering
    \includegraphics[width=1\columnwidth]{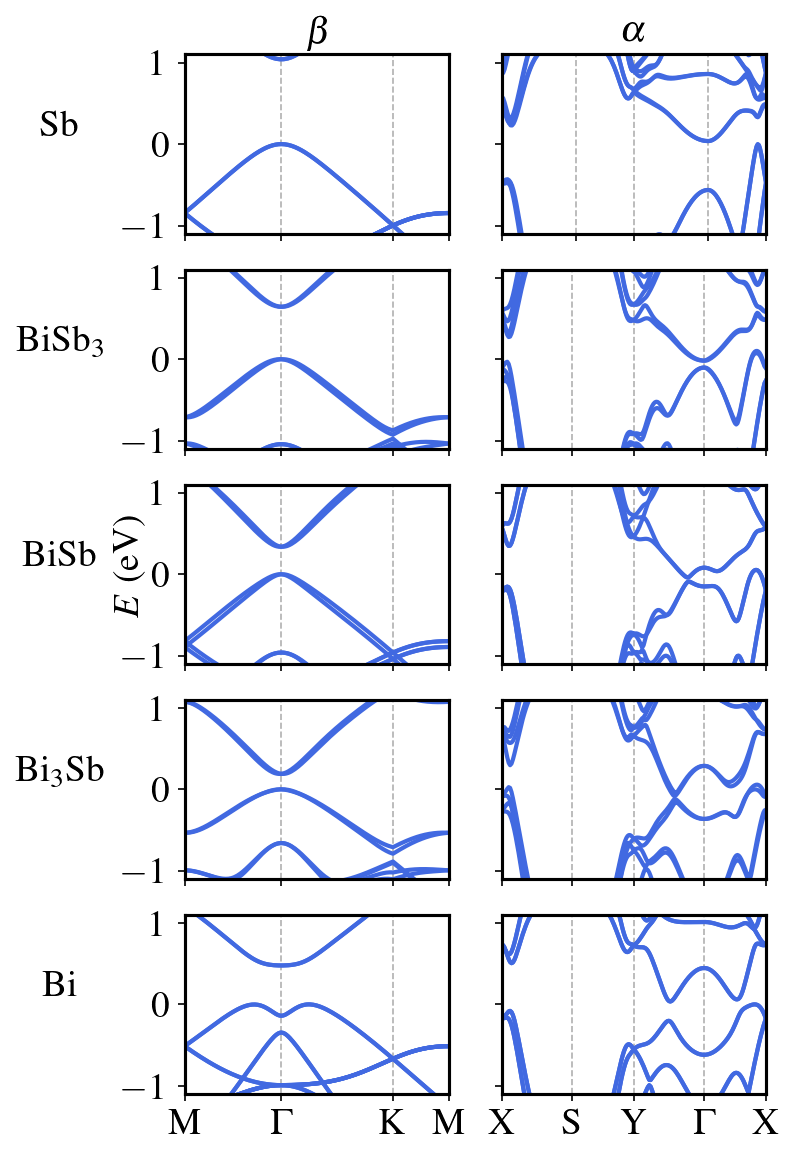}
    \caption{Band structure for five different concentrations of the Bi$_x$Sb$_{1-x}$ alloy, for the $\beta$ structure (left column) and $\alpha$ structure (right column).}
    \label{fig:band_structures}
\end{figure}

We begin applying the previous methodology to obtain the topological phase diagrams of the $\alpha$ and $\beta$ Bi$_{x}$Sb$_{1-x}$ alloys. To do so, we consider nine different stoichiometries corresponding to concentrations $x=i/8$, $i\in\{0,...,8\}$. These ratios allow us to consider the smallest possible unit cells to extract easily the Wilson loop. 
The band structure for some realizations of the alloys for concentrations $x=i/4$, $i\in\{0,...,4\}$ are shown in Fig. \ref{fig:band_structures}, which with the SK parametrization used, we find to be in good agreement with those from DFT \cite{singh2019low, zhou2021atomic, wang2015atomically, bai2022}. As we increase the Bi concentration, we undergo a transition from a trivial system (pure Sb) to a topological one (pure Bi). This can be seen from the fact that as $x$ increases, the gap diminishes until it starts increasing again, but now with a band inversion present. This band inversion appears at the $\Gamma$ point and can be seen in both allotropes. Thus, only from the band structures one could estimate the critical concentrations $x_c$ to be between $0.75 < x_c < 1$ for the $\beta$ crystal and between $0.25<x_c<0.5$ for the $\alpha$ one. However, as we will see these estimates are incorrect; one needs to take into account a bigger crystal and remove artificial periodicities to correctly estimate it.


\begin{figure}
    \centering
    \includegraphics[width=1\columnwidth]{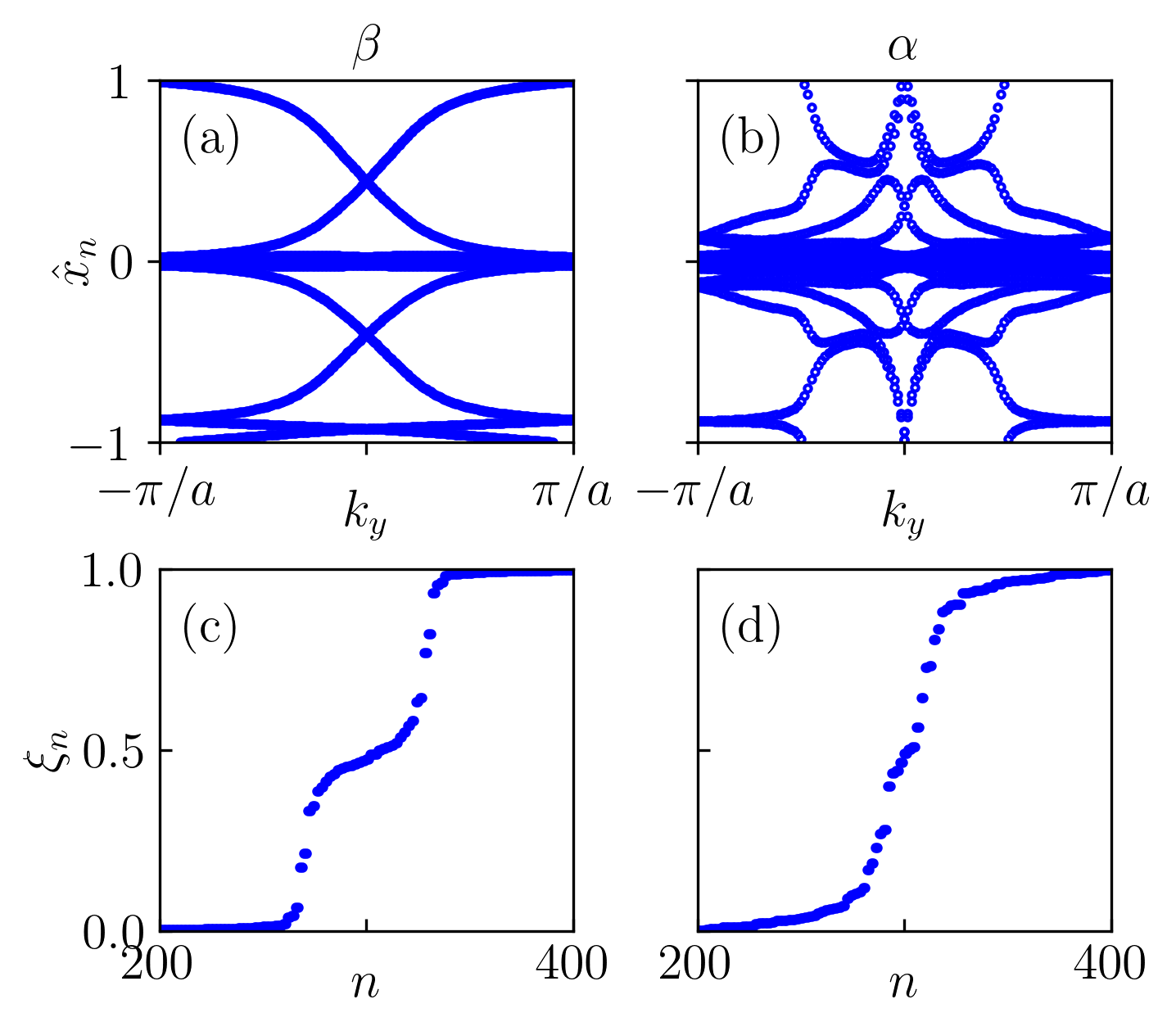}
    \caption{Wannier charge center evolution for (a) $\beta$-Bi$_{0.875}$Sb$_{0.125}$ and (b) $\alpha$-Bi$_{0.875}$Sb$_{0.125}$, both exhibiting topological behaviour. The entanglement spectrum for each system is shown in (c, d) respectively.}
    \label{fig:wilson_loop_entanglement}
\end{figure}

\begin{figure}[t]
    \centering
    \includegraphics[width=1\columnwidth]{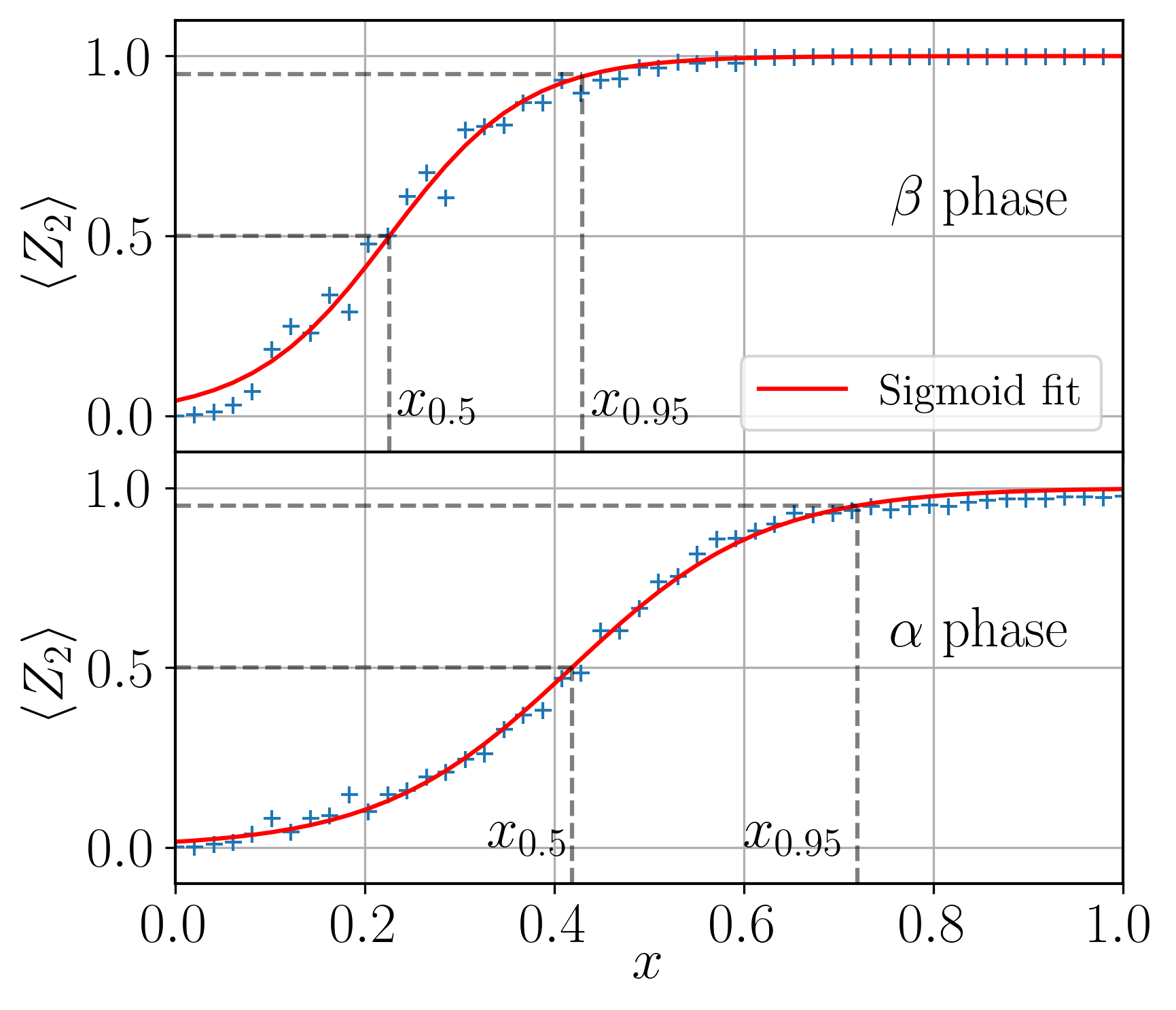}
    \caption{Topological phase diagrams for the (top) beta and (bottom) alpha alloys. The blue points correspond to the prediction of the ANN averaged over $N_s=50$ samples, while the red line is obtained fitting the points to a sigmoid function.}
    \label{fig:crystals_phase_diagrams}
\end{figure}

To obtain the critical concentration in a precise way, we generate training samples for the mentioned nine stoichoimetries, varying the SOC in order to increase the size of the dataset. The resulting datasets are shown in Appendix \ref{sec:training}, Fig. \ref{fig:training_data}. 
An example of the samples used can be seen in Fig. \ref{fig:wilson_loop_entanglement} for both crystals, and for a stoichoimetry corresponding to the topological regime. Finally, following the workflow showed in Fig. \ref{fig:nn_training} we train two ANN, one for each crystal and use them to predict each topological phase diagram. These are shown in Fig. \ref{fig:crystals_phase_diagrams}, where we now plot the average prediction of the neural networks for an intermediate set of concentrations together with a fit to a sigmoid function, $f(x)=(1+e^{-b(x - x_0)})^{-1}$, which typically describes phase transitions. We highlight two different critical concentrations: $x_{0.5}$ which is the standard definition for critical concentrations as it signals that it is more likely to be in the topological region ($p>0.5$). Additionally, we define $x_{0.95}$ in order to establish a critical concentration that indicates a \textit{global} transition of the alloy from trivial to topological. Given how the entanglement spectrum works, it suffices to make the spatial cut across a \textit{locally topological} region to mark the system as topological, while it could be trivial elsewhere. Namely, we take the probability $p$ of the neural network as the probability of the entanglement cut being at a topological region. For this reason, only for $p > 0.95$ we assume that the alloy is globally topological.

\begin{figure}[t]
    \centering
    \includegraphics[width=1\columnwidth]{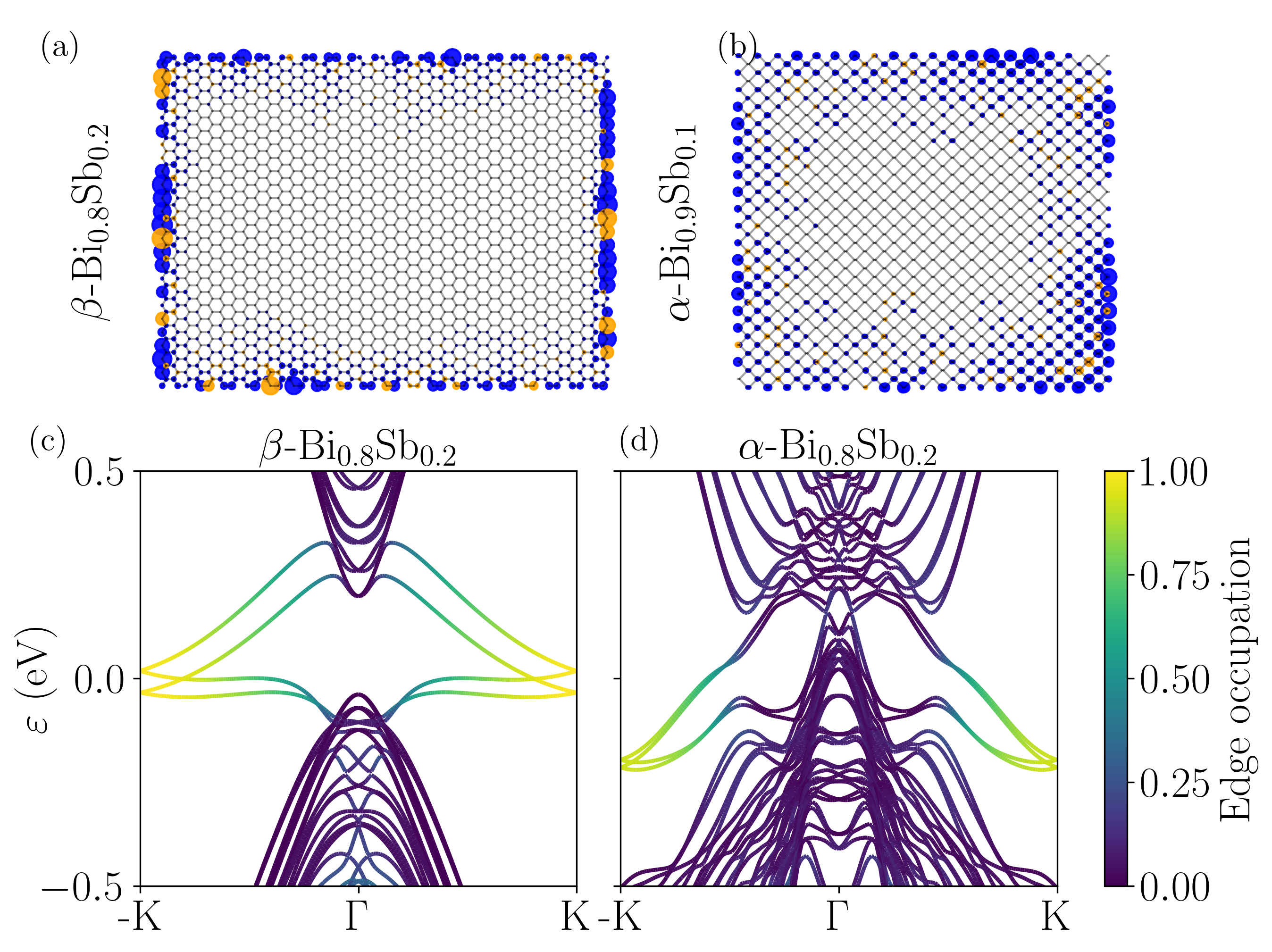}
    \caption{Examples of (a) topological edge state of the $\beta$-Bi$_{0.8}$Sb$_{0.2}$ alloy and (b) edge state of $\alpha$-Bi$_{0.9}$Sb$_{0.1}$. Electronic band structure of a ribbon of (c) $\beta$-Bi$_{0.8}$Sb$_{0.2}$ (50 atoms wide) and (d) $\alpha$-Bi$_{0.8}$Sb$_{0.2}$ (52 atoms wide). The color of the bands represents the probability of the state being localized at the edges of the ribbon.}
    \label{fig:crystals_edge_states_bands}
\end{figure}

\begin{figure}
    \centering
    \includegraphics[width=1\columnwidth]{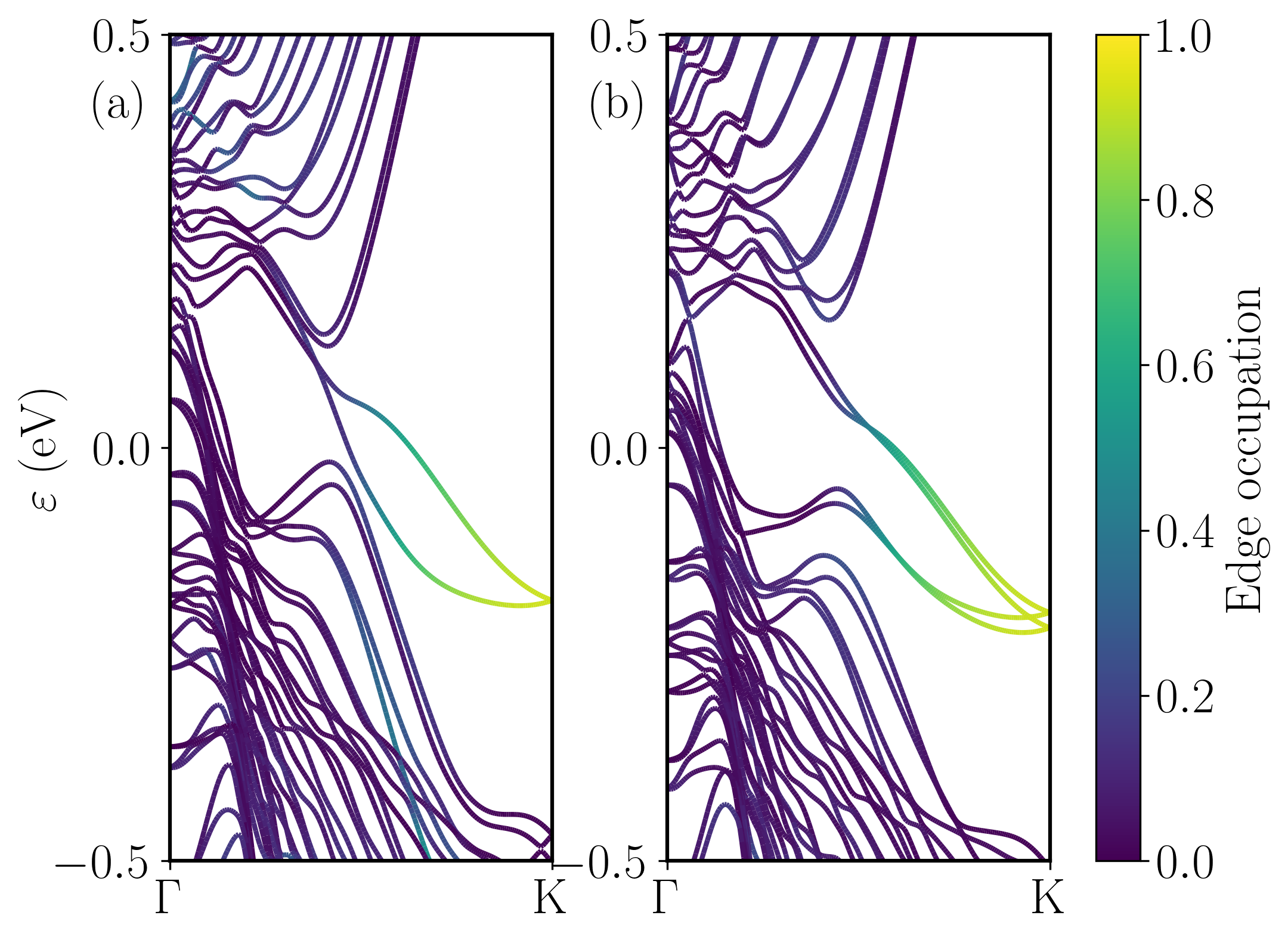}
    \caption{Comparison of the band structure of a ribbon of $\alpha$-Bi$_{0.8}$Sb$_{0.2}$ (52 atoms wide) for (a) topologically trivial sample and (b) non-trivial. }
    \label{fig:alpha_bands_trivial_topological}
\end{figure}

To verify the findings of the ANN, we compute the band structure of a ribbon of both alloys for some concentration beyond $x_{0.95}$. Then, if the system is topological, we should see topological edge bands connecting the valence and conduction bands. In Fig. \ref{fig:crystals_edge_states_bands}(c) we show the bands of a ribbon of $\beta$-Bi$_{0.8}$Sb$_{0.2}$, where each band is colored according to the weight of the wavefunction that is located at the boundaries of the ribbon. Then, we can identify perfectly the topological edge bands. In the case of the $\alpha$ alloy, in Fig. \ref{fig:crystals_edge_states_bands}(d) we show the bands of an instance of the alloy that is topological. In this case, as opposed to the $\beta$ alloy, there is not a gap in the ribbon as a consequence of the bulk band structure, as depicted in Fig. \ref{fig:band_structures}, where there is a direct gap in the system, but not a global gap across the whole BZ. To ensure that the edge bands are indeed topological, we can compare them with those of the trivial case, as shown in Fig. \ref{fig:alpha_bands_trivial_topological}. The SK model we are considering presents trivial edge bands in the trivial regime (for both crystals), which could be confused with the topological ones. From Fig. \ref{fig:alpha_bands_trivial_topological} their distinction becomes clear, as the topological ones still connect valence and conduction bands. 

As an additional check of topological behaviour, we also look for topological edge states in a finite sample of both alloys, again for a concentration beyond the critical one $x_{0.95}$. In Fig. \ref{fig:crystals_edge_states_bands}(a,b) we plot an instance of a topological edge state for the $\beta$-Bi$_{0.8}$Sb$_{0.2}$ and $\alpha$-Bi$_{0.9}$Sb$_{0.1}$ alloys respectively. In both cases we identify the standard behaviour of a topological edge state, with the majority of its weight located at the boundary of the solid. Note that due to the small gap, for the $\beta$ alloy the edge state penetrates into the solid. This is more prominent in the $\alpha$ alloy, which exhibits a longer penetration length.

\subsection{Amorphous Bi$_x$Sb$_{1-x}$}

Following the characterization of the crystalline alloys, we now address the topological properties\ of amorphous Bi$_x$Sb$_{1-x}$, this is, as a function of the disorder strength $\sigma$. In the previous section we established that the concentration of the alloy can be effectively substituted by the SOC $\lambda$, where the hopping parameters only renormalize the critical SOC $\lambda_c$, i.e. where the trivial-topological transition is located. For this reason, we drop the distinction between the two species, and instead fix the hopping parameters and onsite energies, and modify only the SOC strength $\lambda$. Also, given the way the hopping parameters are defined in the amorphous solid (see Eq. \ref{eq:hoppings_amorphous}), the $\beta$ crystal will be the reference structure, as its SK model is simpler than the one for the $\alpha$ crystal (namely, the $\beta$ crystal can be described in the crystalline limit of the amorphous model, while this is not possible for the $\alpha$ crystal due to its hopping parameters to different neighbours).

\begin{figure}
    \centering
    \includegraphics[width=1\linewidth]{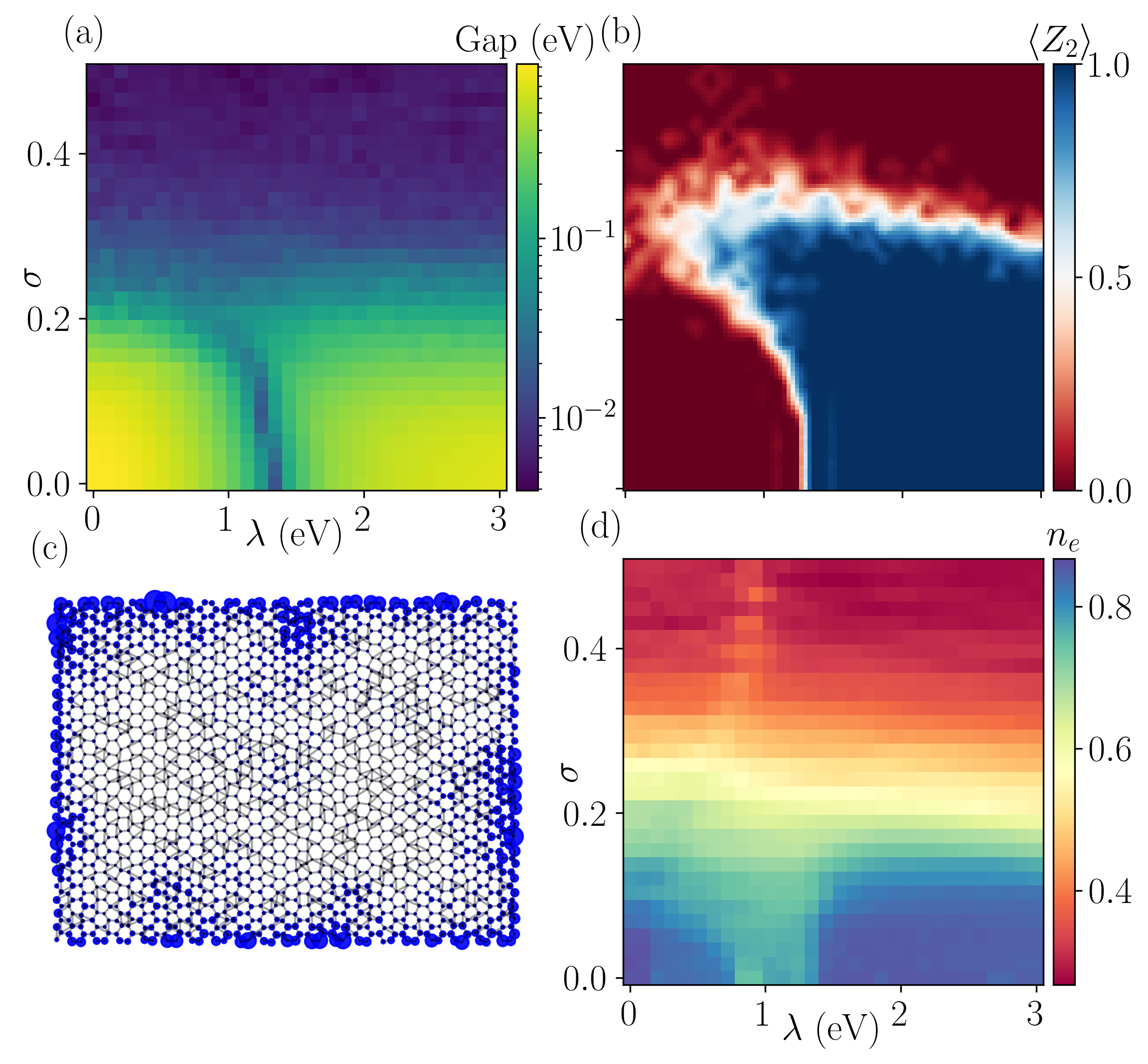}
    \caption{Diagrams for the amorphous alloy as a function of the SOC strength $\lambda$ and the disorder strength $\sigma$ for (a) gap, for a system with $N_c=16^2$ unit cells and averaged over 10 samples, and (b) $\mathbb{Z}_2$ invariant (averaged over 20 samples). (c) Topological edge state for the amorphous solid at $\sigma=0.25$, $\lambda=2.5$ eV. (d) Average edge occupation of the lowest $21$ state energy states as a function of $\sigma$ and $\lambda$, with $N_s=10$ samples and $N_c=20\times 15$ unit cells. At high disorder, there is a decrease in the edge occupation of the lowest states for all values of $\lambda$, possibly signaling a global topological phase transition to a trivial state. }
    \label{fig:gap_topo_diagrams_amorphous}
\end{figure}

A topological phase transition from a topological to a trivial insulator can only take place following a closing of the gap. For this reason we begin the characterization of the amorphous model obtaining the gap of the system to identify potential transitions. The gap as a function of both disorder and SOC strength is shown in Fig. \ref{fig:gap_topo_diagrams_amorphous}(a). Most notably, we distinguish two different regions: at low disorder ($\sigma < 0.3$) there is a closing of the gap that moves with disorder, i.e. $\lambda_c=\lambda_c(\sigma)$, potentially showing a topological phase transition. At high disorder ($\sigma > 0.3$) the system becomes gapless $\forall \lambda$. This region can correspond potentially to four different phases: either an Anderson insulator or a metal, which could then be trivial or topological \cite{groth2009theory, li2009topological}.

To establish the topological nature of all three regions, we resort again to the ANN. We follow the same procedure as before: we generate the dataset to train a new neural network, which consists of pairs of entanglement spectrum plus topological invariant determined via the Wilson loop. In this case, we want to obtain the topological phase diagram as a function of both disorder and SOC. For this reason, we generate training samples at zero and finite disorder, for all the SOC values considered ($\lambda\in[0,3]$). The distribution of the training data can be seen in appendix \ref{sec:training}, in Fig. \ref{fig:training_data}. Once the ANN is trained, we use it to predict the topological phase diagram of the amorphous solid, shown in Fig. \ref{fig:gap_topo_diagrams_amorphous}(b). As we guessed, at low disorder there is a trivial to topological transition. Remarkably, the critical strength $\lambda_c$ depends on the disorder. This implies that for fixed SOC, say $\lambda=1$, if we increase disorder we undergo a transition from trivial to topological. In other words, it is an amorphization-induced topological transition. At high disorder we find the opposite behaviour: for high SOC ($\lambda > 1.3$) disorder destroys the topological phase, transitioning to a gapless trivial region. For low SOC ($\lambda < 1.3$), the system was already in a trivial region, so it does not change in that regard, although it becomes gapless.

\begin{figure}
    \centering
    \includegraphics[width=1\linewidth]{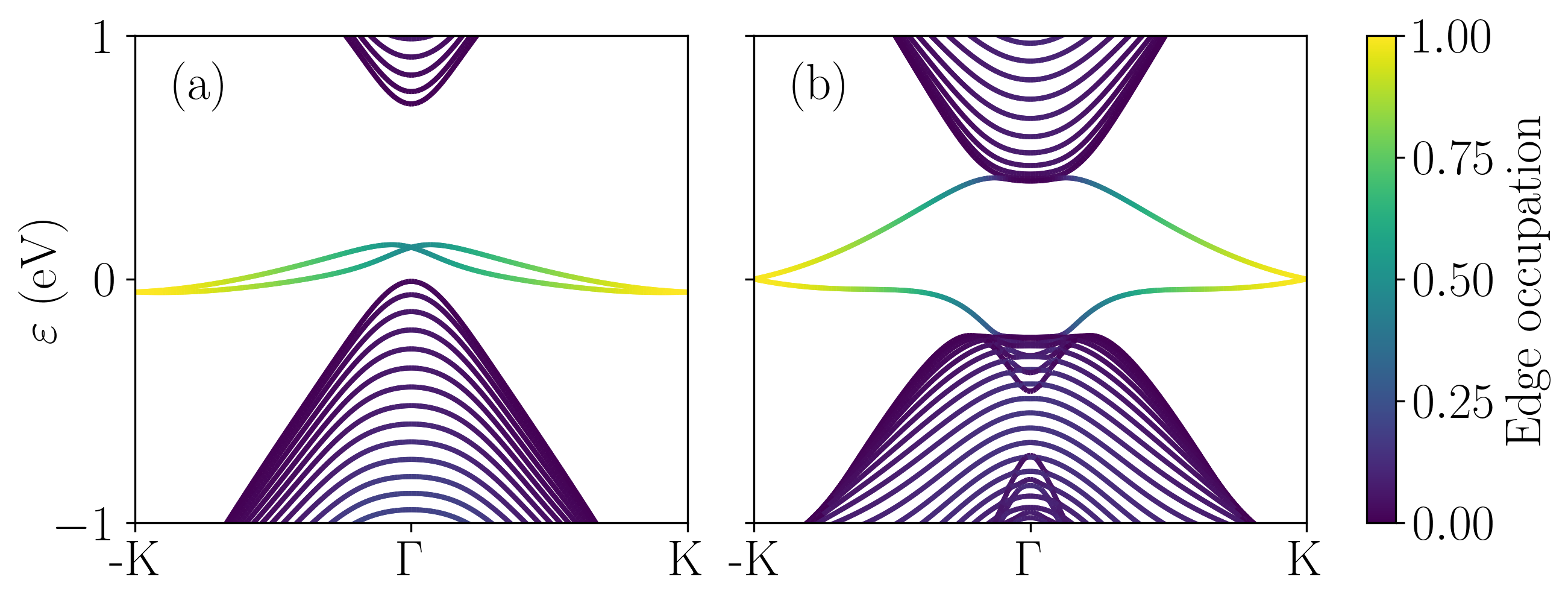}
    \caption{Bands of a $\beta$ (zigzag) ribbon with a width of 50 atoms using the parameters of the amorphous alloy, for two different SOC values: (a) $\lambda=0.5$ eV, which exhibits trivial edge states, and (b) $\lambda=2.5$ eV, showing the topological edge bands. }
    \label{fig:beta_ribbon_bands}
\end{figure}

To verify the results of the ANN, we look for signatures of the topological edge states. First, we evaluate the average edge occupation $\overline{\braket{n_e}}=1/N\sum_n^N\sum_{i\in\partial\Omega}\braket{c^{\dagger}_ic_i}_n$ for the first $N$ states closer to the Fermi energy, where $\partial\Omega$ denotes the boundary of the solid $\Omega$. This quantity gives a direct measure of the degree of localization of the states, which for topological edge states should be close to 1. The average edge occupation is shown in Fig. \ref{fig:gap_topo_diagrams_amorphous}(d). There, we observe that both trivial and topological regions at low disorder exhibit high edge occupations. For the trivial region, this is due to the presence of trivial edge states, whereas in the topological is due to the topological edge states, as depicted in Fig. \ref{fig:beta_ribbon_bands}. The transition between the regions is characterized by a lower edge occupation, as it is expected from the increasingly higher bulk component of the edge states as the gap closes. At high disorder, the edge occupation drops significantly for all values of $\lambda$. This is to be expected due to the appearance of localized bulk states coming from the disorder, which will reduce the average even if there are proper topological edge states present. 

Next, we look explicitly for the presence of topological edge states. One instance of such a state is shown in Fig. \ref{fig:gap_topo_diagrams_amorphous}(c), at the limit of the topological region with $\sigma=0.25$. In this case, the edge state shown already shows some bulk component, which will increase as disorder increases (gap decreases). Looking for edge states at high disorder ($\sigma > 0.3$), we are unable to find a well-defined edge state since all of them exhibit strong bulk components and in general do not wrap the solid as one would expect.

\begin{figure*}
    \centering
    \includegraphics[width=1\linewidth]{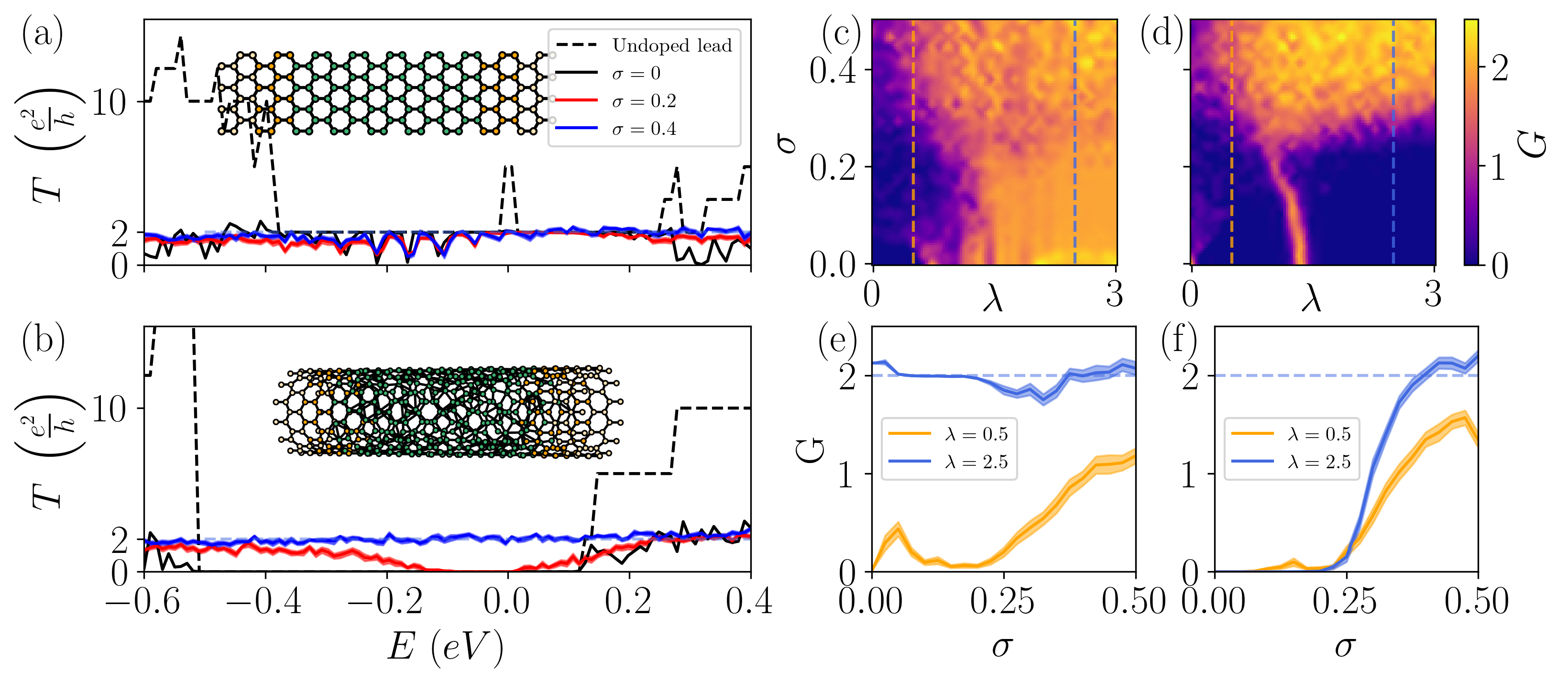}
    \caption{(a,b) Transmission as a function of disorder of the Bi$_x$Sb$_{1-x}$ alloy for a nanoribbon (OBC) and a nanotube (PBC) respectively, for $\lambda=2.5$ and averaged over 10 samples. (c) Conductance diagrams as a function of SOC strength $\lambda$ and disorder strength $\sigma$ for OBC and (d) PBC respectively, averaged over 10 samples. The dashed lines correspond to the cuts that are represented on subplots (e,f), which show the conductance as a function of disorder for different $\lambda$ for OBC and PBC respectively, averaged over 50 samples (shaded areas denote standard error of the mean). All calculations have been done with system size $W=16$, $L=9$.}
    \label{fig:transport}
\end{figure*}

\begin{figure}[h]
    \centering
    \includegraphics[width=1\columnwidth]{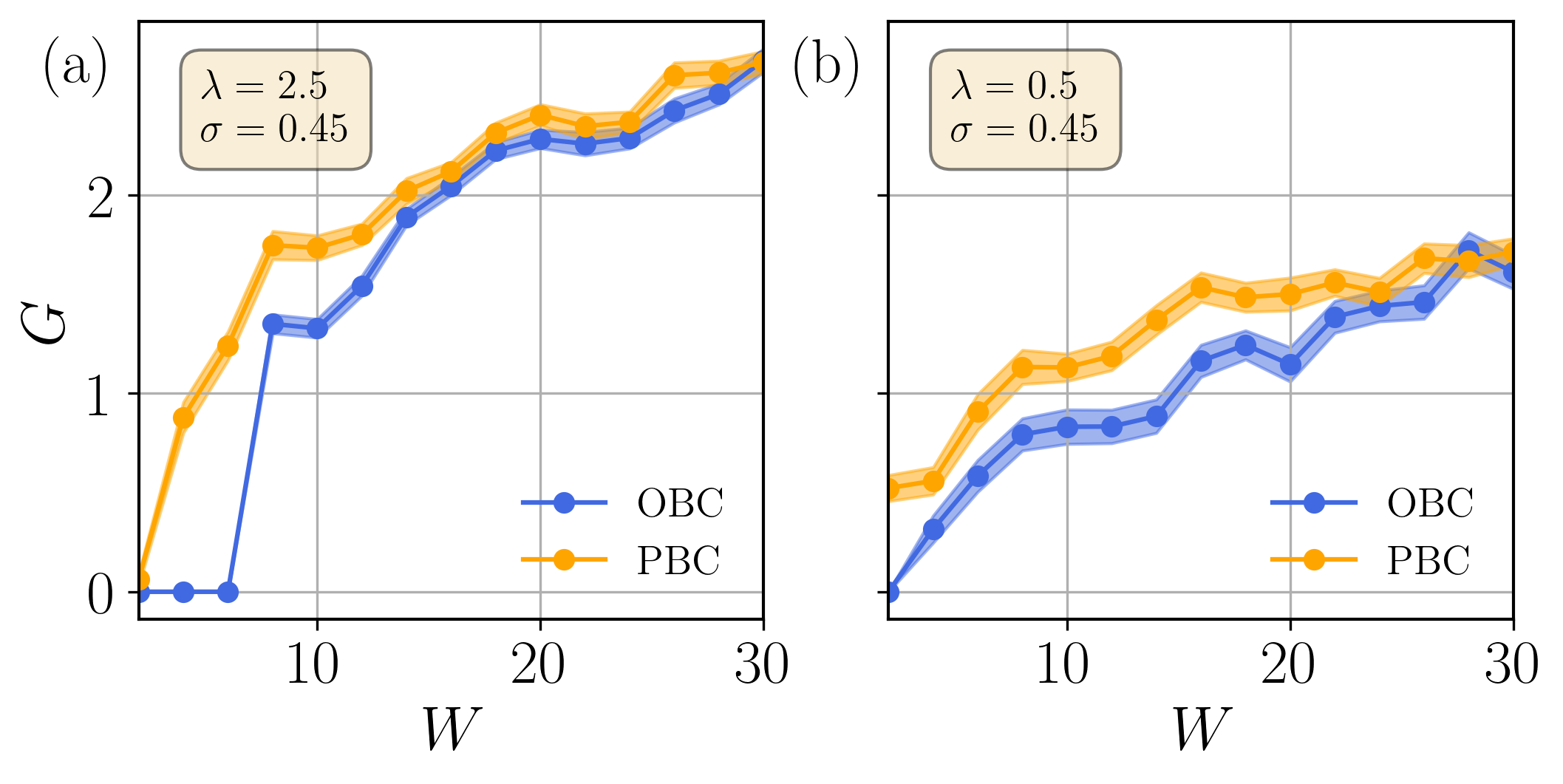}
    \caption{Scaling of the conductances for high disorder $\sigma=0.45$ as a function of $W$ ($=L$) for (a) $\lambda=2.5$ and (b) $\lambda=0.5$ for both OBC and PBC. We have chosen pairs $(W,L)$ such that the aspect ratio of the ribbon is closest to 1. All points are averaged over $N_s=50$ samples, and the shaded areas denote the standard error of the mean $\sigma_{\mu}=\sigma/\sqrt{N_s}$.}
    \label{fig:scaling}
\end{figure}

Finally, to ensure that the topological edge states do not play any role in the high disorder region, we set up electronic transport calculations. The aim is two-fold: to establish whether there are topological edge states contributing to the conductance and to determine the transport nature of the bulk system, this is, insulating or metallic. For this, we consider two different setups: with open boundary conditions (i.e. transport along a ribbon), and with periodic boundary conditions (transport along a nanotube). Examples of both transport setups are shown in Fig. \ref{fig:transport}(a,b) respectively. The leads are taken to be of the same material as the sample, but doped to ensure that they are always metallic irrespective of the sample behaviour. For more details on the setup, we refer to appendix \ref{app:transport}.

In the first place, we determine the transmission with both OBC and PBC as a function of disorder for a topological sample with $\lambda=2.5$, shown in Fig. \ref{fig:transport}(a,b). For the nanoribbon, we observe that for all values of disorder the transmission at the Fermi level is approximately two (the Fermi level is set at zero). On the contrary, for the nanotube at low disorder the sample is insulating as expected, but at high disorder its transmission also increases to two, meaning that the bulk states contribute to the conduction. This is further verified with the conductance diagrams in Fig. \ref{fig:transport}(c,d), also for OBC and PBC respectively. For OBC at low disorder, we can identify the topological region as it exhibits a quantized conductance $G=G_0=2e^2/h$. At high disorder the conductance starts fluctuating and increases beyond the quantum of condutance $G_0$. For the PBC diagram, we observe the same: the trivial and topological regions at low disorder are insulating as expected, but in the high disorder region the sample becomes metallic. The spurious conductances seen at low disorder in both cases are attributed to the trivial edge states in the OBC case, and to the proximity to the valence band in the PBC case (see Fig. \ref{fig:transport}(e,f)).

To conclude the metallic nature of the system, we perform an scaling analysis in the spirit of the scaling theory of localization \cite{abrahams1979scaling}. Namely, as we increase simultaneously the width $W$ and the length $L$ of the sample and the leads, the bulk system must show one of two different behaviours: either its conductance increases with the sample size (metallic, more channels available for transport), or the conductance decays to zero (insulating, due to exponentially localized states). We examine the scaling of the system for both OBC and PBC at high disorder, $\sigma=0.45$. From the diagrams in Fig. \ref{fig:transport}(c,d) we also observe a difference in the conductance at low SOC ($\lambda =0.5$) and high SOC ($\lambda=2.5$). For this reason we address the scaling for both values, to detect any possible difference between these two metallic regions. The scaling analysis is illustrated in Fig. \ref{fig:scaling}. For both values of $\lambda$ we observe an increase of the conductance with the system size, in principle indicating a metallic scaling. Regarding the difference between OBC and PBC, while the conductances show some differences for small system sizes, these discrepancies decrease as the system size increases, hinting that there is not any contribution from topological edge states in the OBC case. In the particular case of $\lambda=0.5$, it appears the PBC conductance could be plateauing, which begs the question of what happens for bigger system sizes, namely, if it would still be possible that the PBC conductance drops to an insulating state whereas the OBC one plateus due to topological edge states. Unfortunately, those system sizes are beyond our reach for the present study. With this we conclude that the system undergoes a insulator-metal transition with disorder, which was previously observed in the context of Anderson insulators and dubbed inverse Anderson insulators \cite{goda2006inverse, zuo2024topological}.

\section{Conclusions}

Using the entanglement spectrum as a proxy for the $\mathbb{Z}_2$ topological invariant in conjunction with artificial neural networks, we were able to determine the topological phase diagram of the crystalline Bi$_x$Sb$_{1-x}$ alloys in their $\beta$ and $\alpha$ forms, estimating the critical concentrations. Applying this same methodology to the amorphous alloy, we observe that for low disorder strength, this is in the gapped region, the structural disorder renormalizes the critical SOC, resulting in trivial to topological transitions. This result was already observed in SK models of stanane \cite{wang2022}, and we conjecture that it extends to all families of topological insulators that accept the same description. Additionally, we go beyond previous works to study the high disorder region, i.e. where the system becomes gapless. With the ANN predicting a global transition to a trivial region, we resort to electronic transport calculations. Remarkably, we observe that the system undergoes a transition from an insulator to a metal with disorder, although it is possible that the metallic phase exhibits different regimes. While in principle we have discarded any topological behaviour in the $\mathbb{Z}_2$ sense, it would be interesting to explore other forms of topology, namely HOTIs, as other forms of Bi exhibit, and in general a more detailed study of the scaling would benefit the characterization of the metallic phase to ensure the distinction between trivial and topological. We also highlight the performance of the neural network, which was able to predict the phase diagram independently of the gap of the system.\\

\begin{acknowledgments}
A. J. Ur\'ia-\'Alvarez is thankful to A. G. Grushin for the helpful discussions. The authors acknowledge financial support from Spanish MICINN (Grant Nos. PID2019-109539GB-C43, TED2021-131323B-I00 \& PID2022-141712NB-C21), María de Maeztu Program for Units of Excellence in R\&D (GrantNo.CEX2018-000805-M), Comunidad Autónoma de Madrid through the Nanomag COST-CM Program (GrantNo.S2018/NMT-4321) and (MAD2D-CM)-UAM7 project funded by the Comunidad de Madrid, by the Recovery, Transformation and Resilience Plan from Spain, and by NextGenerationEU from the European Union; Generalitat Valenciana through Programa Prometeo (2021/017), Centro de Computación Científica of the Universidad Autónoma de Madrid and Red Española de Supercomputación.

\end{acknowledgments}

\begin{appendix}

\section{Slater-Koster parametrizations}
\label{sec:sk_parameters}

As it was specified in the main text, we use the parameters of \cite{liu1995} to parametrize the alloys for both crystals. Taking into account the mixing of the parameters that is performed whenever there is a hopping between two different species, we only show the parameters for both lattices for the Bi atoms (the same applies to the Sb atoms).

\renewcommand{\arraystretch}{1.5}
\begin{table}[h]
    \centering
    \begin{tabular}{ccccc}
    \hline
    \hline
       $n$-th nn  & $V_{ss\sigma}$ & $V_{sp\sigma}$ & $V_{pp\sigma}$ & $V_{pp\pi}$ \\
    \hline
    1nn & -0.608 & 1.32 & 1.854 & -0.6 \\
    \hline
    \hline
    \end{tabular}
    \caption{Hopping amplitudes up to the $n$-th next neighbour ($n$-th nn) for the $\beta$ crystal. We show the specific values for Bi; the same holds for the description of Sb with the corresponding hopping parameters. Units are eV.}
    \label{tab:sk_beta}
\end{table}

\begin{table}[h]
    \centering
    \begin{tabular}{ccccc}
    \hline
    \hline
       $n$-th nn  & $V_{ss\sigma}$ & $V_{sp\sigma}$ & $V_{pp\sigma}$ & $V_{pp\pi}$ \\
    \hline
    1nn & -0.608 & 1.32  & 1.854 & -0.6   \\
    2nn & -0.453 & 0.984 & 1.382 & -0.447 \\
    3nn &   0    &   0   &   0   &    0   \\
    4nn &   0    &   0   & 0.156 &    0   \\
    \hline
    \hline
    \end{tabular}
    \caption{Hopping amplitudes (in eV) up to the $n$-th next neighbour ($n$-th nn) for the $\alpha$ crystal. We show the specific values for Bi; the same holds for the description of Sb with the corresponding hopping parameters.}
    \label{tab:sk_alpha}
\end{table}

The onsite energies correspond directly to those from \cite{liu1995} so we do not show them. For the $\beta$ crystal, we only use the hopping parameters up to first neighbours (even though the original model also provides them up to second), to use the model as the starting point for the amorphous case. For the $\alpha$ alloy, we needed more complex hopping parameters to reproduce the band structures. We introduce hopping parameters to second neighbours, obtained from the hopping parameters to 1-nn but scaled with $t(r)=t_0(r_0/r)^{1.5}$, where $t_0$, $r_0$ are the reference hopping and bond length. For third and fourth nn hopping parameters, we used those to 2-nn from the original model such that they gave similar bands to DFT. All three models are implemented using the \texttt{tightbinder} code \cite{uria2024tightbinder}.

\section{Calculation of the entanglement spectrum with the kernel polynomial method}

We used the kernel polynomial method (KPM) \cite{RevModPhys.78.275} to generate part of the training dataset for the neural network as it allows to obtain the correlation matrix $C_{ij}$ without having to diagonalize the Hamiltonian. We follow the approach taken in previous works \cite{varjas2020, carvalho2018} where the projector over the Fermi sea can be written in terms of a function of the Hamiltonian $H$. This allows to rewrite the correlation matrix directly in terms of this function of $H$, which can then be expanded in powers of $H$ according to the KPM.
\begin{align}
    \nonumber C_{ij}&=\braket{\Psi|c^{\dagger}_ic_j|\Psi}=\sum_{n }^{E_n\leq E_F}\braket{n|i}\braket{j|n} \\ 
    &=\braket{j|\left(\sum_n^{E_n\leq E_F}\ket{n}\bra{n}\right)|i} = \braket{j|\theta(E_F - H)|i}
\end{align}
where $E_F$ is the Fermi energy, and $\theta(E)$ is the Heaviside step function. Thus, we identify the function of the Hamiltonian $P(E, H)=\theta(E - H)$. Normalizing the Hamiltonian such that its eigenvalues lie in the interval $[-1, 1]$, we can expand this function in Chebyshev polynomials:
\begin{equation}
    P(E,H) = \sum^M_{m=0}g_m\mu_m(E)T_m(H)
\end{equation}
where $g_m$ are the Jackson kernel coefficients, $\mu_m(E)$ are the moments and $T_m(H)$ are the Chebyshev polynomials, defined in terms of a recursive relation:
\begin{align}
    \nonumber T_0(H) &= I \\
    T_1(H) &= H \\
    \nonumber T_m(H) &= 2HT_m(H) - T_{m-1}(H)
\end{align}
The moments $\mu_m(E)$ can be evaluated knowing that $P(E,H)=\theta(E - H)$ and read \cite{varjas2020}:
\begin{align}
    \mu_m(E) &= \frac{2}{\pi}\frac{1}{1+\delta_{m0}}\int_{-1}^1\frac{P(E,x)T_m(x)}{\sqrt{1 - x^2}}dx \\ 
    &=\left\{
    \begin{array}{cc}
         1 - \frac{1}{\pi}\text{arccos(E)} & m=0 \\
        \frac{-2}{m\pi}\sin[{m\ \text{arccos}(E)}] & m\neq 0 
    \end{array}\right.
\end{align}
Finally, taking matrix elements of $P(E,H)$ in the orbitals basis $\{\ket{i}\}$ we obtain the correlation matrix $C_{ij}$. The eigenvalues of this matrix form the entanglement spectrum used to train the neural network and predict the topological nature of the alloys. Thus, given that we still need to diagonalize the correlation matrix, the speedup provided by the KPM is restricted by this diagonalization of a matrix of dimension $N/n$, where $N$ is the dimension of the complete system and $n$ is the reduction factor coming from the restriction to a partition of the system. Therefore, time complexity is still $\mathcal{O}(N^3)$, although in practice it leads to faster calculations. 

\section{Training data and neural networks}
\label{sec:training}

\begin{figure}
    \centering
    \includegraphics[width=1\linewidth]{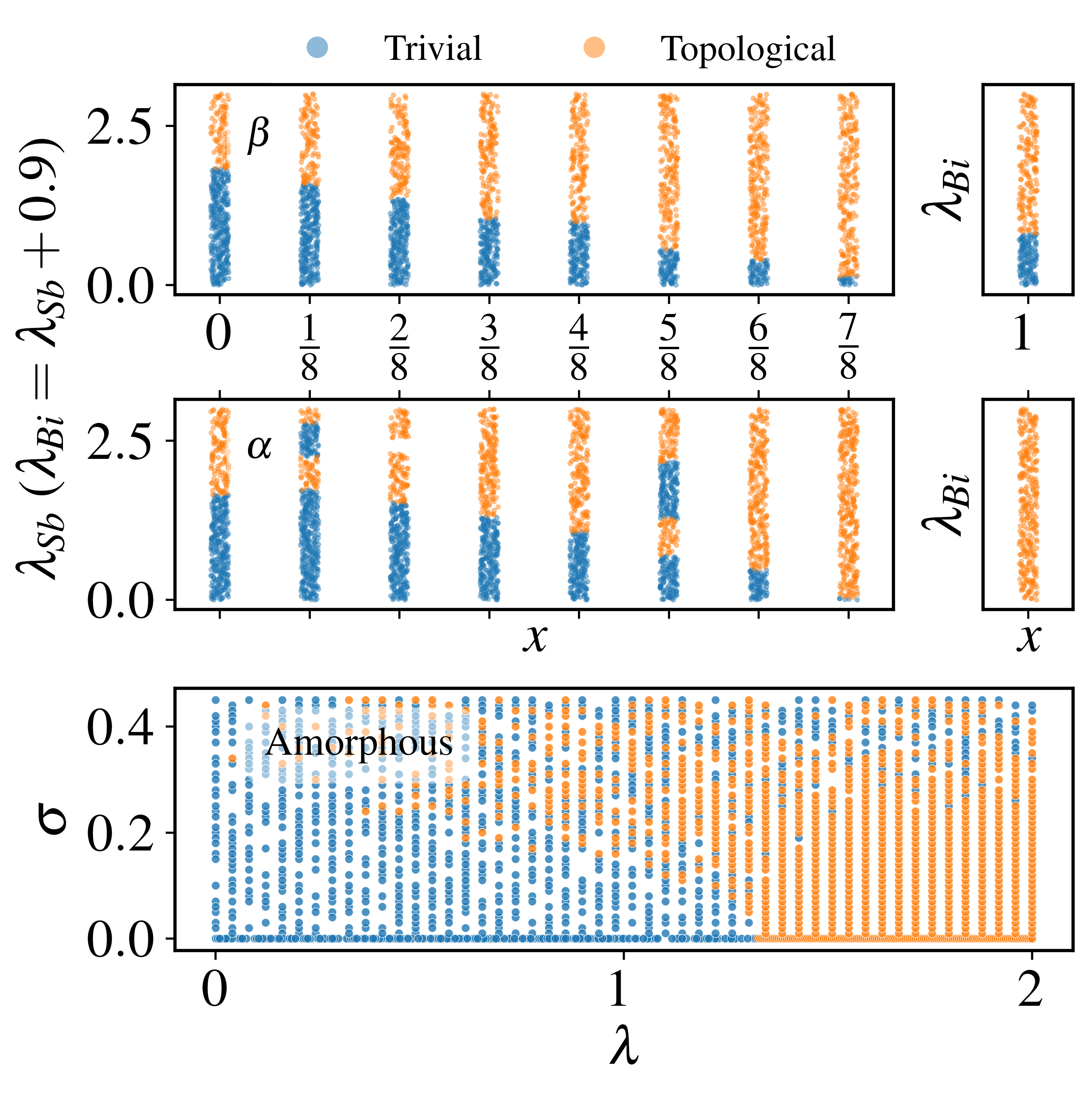}
    \caption{Datasets used in the training of the neural networks. (Top) Dataset for the $\beta$ alloy, (middle) dataset for the $\alpha$ alloy and (bottom) dataset for the amorphous alloy. In the crystalline cases, the samples are represented as a function of the SOC of Sb, $\lambda_{\text{Sb}}$; the SOC for Bi is always defined as  $\lambda_{\text{Bi}}=\lambda_{\text{Sb}}+0.9$. For $x=1$ there is no Sb, so the points are represented directly as a function of $\lambda_{\text{Bi}}$.}
    \label{fig:training_data}
\end{figure}

The datasets used to train and test the three neural networks are represented in Fig. \ref{fig:training_data}. As explained in the main text, for the crystalline alloys we generate data considering a finite set of concentrations and changing the value of the SOC. The datasets for the $\beta$ and $\alpha$ alloy contain a total of 4500 samples each, corresponding to 500 per concentration. After balancing the datasets to ensure the same number of samples per category, we get approximately a total of 2900 samples for the $\beta$ alloy, and 3300 samples for the $\alpha$ alloy.

For the amorphous alloy, we take directly the SOC as an effective concentration parameter, and change the disorder strength to generate samples for both crystalline and disordered cases. In this case, we generated a total of 2600 samples, which after balancing resulted in 2100 samples.

We used two different architectures for the neural networks, one for the crystalline case and a different one for the amorphous case.
These architectures are shown in Table \ref{tab:arch_crystalline} and Table \ref{tab:arch_amorphous} respectively. The main idea was to use a convolutional layers to extract the main features of the entanglement spectrum, followed by dense layers to perform the interpolation. In the case of the amorphous network, it seemed to benefit from the inclusion of additional convolutional and dense layers, increasing its accuracy. All layers use a ReLU activation function, except for the last layer which uses a sigmoid.

In training, we used a train-test split of $10\%$. For the $\beta$ network, we achieve an accuracy on the test set of $99\%$, while for the $\alpha$ network we achieve $85\%$. Both networks were trained for 50 epochs, with a learning rate $\alpha=10^{-4}$ with the ADAM optimizer. The amorphous network was trained for 200 epochs, also with $\alpha=10^{-4}$, achieving an accuracy on the test set of $85 \%$.

\renewcommand{\arraystretch}{1.2}
\begin{table}[h]
    \centering
    \begin{tabular}{ccc}
        \hline
        \hline
        Layer type & Kernel & Output shape \\
        \hline
        1d convolutional & 16 &  $32\times200\times1$ \\
        Max-pool & 2 & $32\times100\times1$ \\
        Dense & - & 1000 \\
        Dense & - & 1 \\
        \hline
        \hline
    \end{tabular}
    \caption{Architecture of the ANN used to predict the topological phase diagram of the $\beta$ and $\alpha$ alloys.}
    \label{tab:arch_crystalline}
\end{table}

\begin{table}[h]
    \centering
    \begin{tabular}{ccc}
        \hline
        \hline
        Layer type & Kernel & Output shape \\
        \hline
        1d convolutional & 32 &  $32\times200\times1$ \\
        Max-pool & 2 & $32\times100\times1$ \\
        1d convolutional & 64 &  $64\times100\times1$ \\
        Max-pool & 2 & $64\times50\times1$ \\
        1d convolutional & 128 &  $128\times50\times1$ \\
        Max-pool & 2 & $128\times25\times1$ \\
        Dense & - & 256 \\
        Dense & - & 10 \\
        Dropout ($p=0.1$) & - & - \\
        Dense & - & 10 \\
        Dense & - & 1 \\
        \hline
        \hline
    \end{tabular}
    \caption{Architecture of the ANN used to determine the phase diagram of the amorphous alloy.}
    \label{tab:arch_amorphous}
\end{table}

\section{Transport setup}
\label{app:transport}

\begin{figure}[h]
    \centering
    \includegraphics[width=1\linewidth]{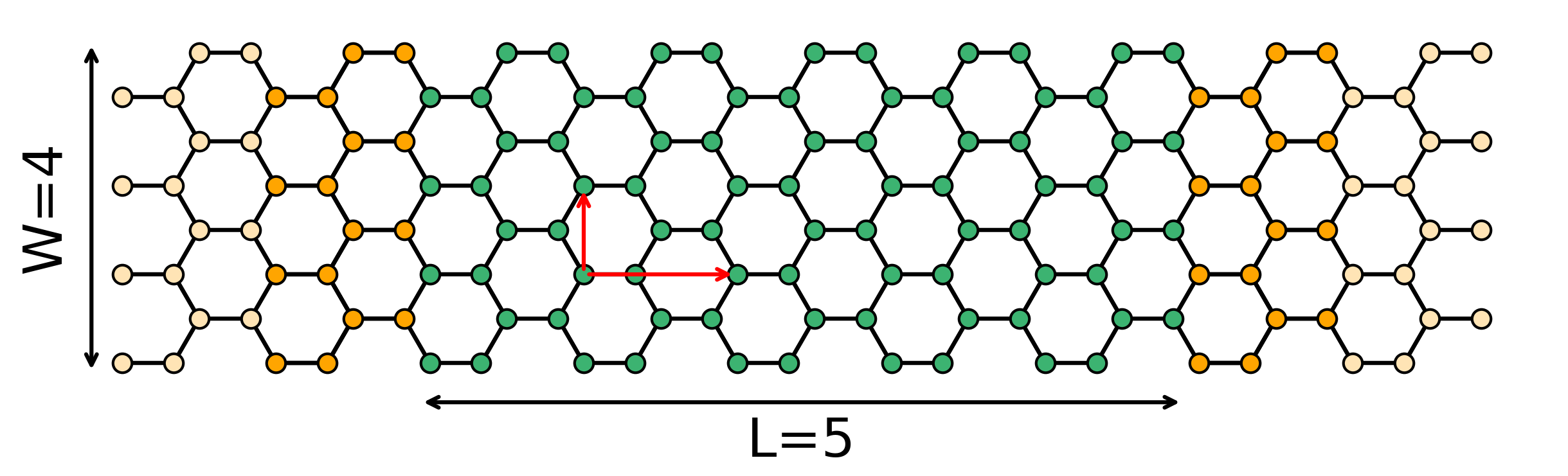}
    \caption{Transport setup for an OBC calculation. This setup corresponds to a sample (green atoms) with $L=5$ and $W=4$ unit cells. The orange and yellow atoms denote the leads attached to the sample. The red arrows indicate the unit cell of the ribbon. The device $D$ refers to the sample together with the first unit cell of the lead (green and orange atoms).}
    \label{fig:transport_setup_obc}
\end{figure}

\begin{figure}
    \centering
    \includegraphics[width=0.9\linewidth]{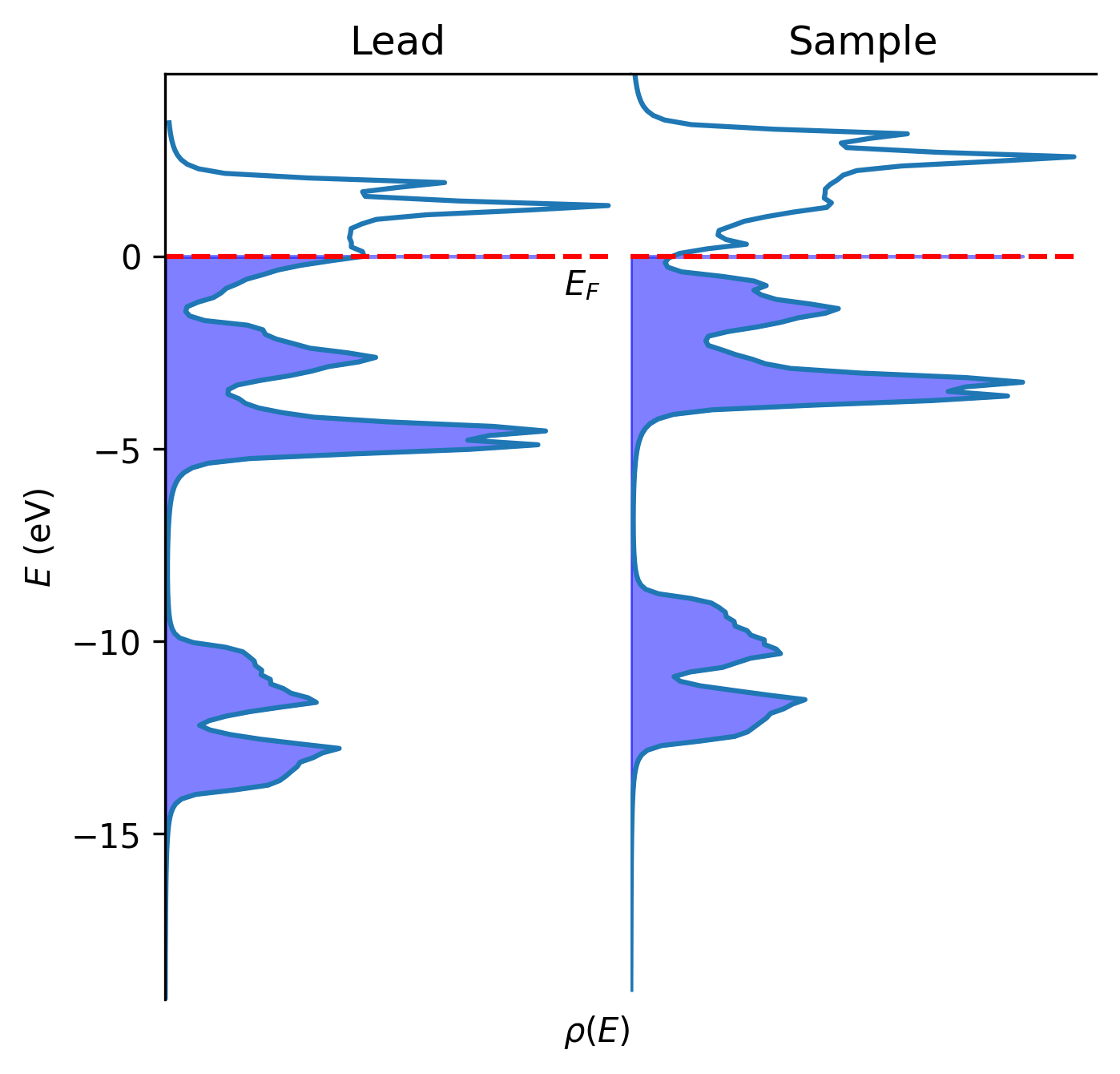}
    \caption{Density of states of (left) leads, with an applied gate voltage to impose an specific doping and (right) sample.}
    \label{fig:dos_sample_leads}
\end{figure}

We use Landauer's formalism to determine the electronic transport properties of a sample of the amorphous alloy \cite{datta1997electronic}. The conductance of the sample is given by Landauer's formula \cite{landauer1957spatial}:
\begin{equation}
    G = \frac{e^2}{h}T(E_F)
\end{equation}
where $E_F$ is the Fermi energy of the sample, and $T$ is the transmission, which is computed with the Caroli formula \cite{caroli1971direct}:
\begin{equation}
    T(E) = \text{Tr}\left[\Gamma_L(E)G_D^-(E)\Gamma_R(E)G_D^+(E)\right]
\end{equation}
where $\Gamma_{L/R}$ denotes the coupling of the leads with the device (sample plus unit cell of the lead, see Fig. \ref{fig:transport_setup_obc}), given by $\Gamma_{L/R}=i(\Sigma_{L/R}^+ - \Sigma^-_{L/R})$ with $\Sigma_{L/R}^{+/-}$ being the selfenergy of lead $L/R$, and $G_D$ is the Green's function of the device. For the definition of these quantities we refer to previous works \cite{jacob2011critical}.

The conductance is obtained at the Fermi level of the sample. To ensure that the leads are always metallic and provide the same current for different instances of disorder, we consider a gate potential applied to the leads to ensure constant charge across calculations. The gate potential for a fixed charge in the leads is obtained integrating the density of states of the leads:
\begin{equation}
    \int_{-\infty}^{E_F}\rho_{L/R}(E - V)dE = N_e
\end{equation}
where $N_e$ is the desired charge. In all calculations, we have set $N_e=N + N_a/2$, where $N$ is the number of electrons of the lead at charge neutrality, and $N_a$ is the number of atoms of the lead. This level of doping ensures that the Fermi level of the sample is aligned with the middle of the conduction band of the lead, resulting in constant transport across the leads. The density of states of both the sample and the lead are shown in Fig. \ref{fig:dos_sample_leads}, with the gate voltage already applied to the leads. The transport calculations were done with the \texttt{tightbinder} library \cite{uria2024tightbinder}.

\end{appendix}

\bibliography{biblio}

\end{document}